\documentclass[pre,a4paper,showpacs]{revtex4-1}
\usepackage{graphicx}
\usepackage{dcolumn}
\usepackage{amsmath}
\usepackage{array}
\usepackage{bm}
\usepackage{amssymb}
\usepackage{amsfonts}

\begin{document}

\title{Lagrange-mesh calculations in momentum space}

\author{Gwendolyn Lacroix}
\thanks{F.R.S.-FNRS Research Fellow}
\email[E-mail: ]{gwendolyn.lacroix@umons.ac.be}
\author{Claude \surname{Semay}}
\thanks{F.R.S.-FNRS Senior Research Associate}
\email[E-mail: ]{claude.semay@umons.ac.be}
\author{Fabien \surname{Buisseret}}
\email[E-mail: ]{fabien.buisseret@umons.ac.be}
\altaffiliation{Haute \'Ecole Louvain en Hainaut (HELHa), Chauss\'ee de Binche 159, 7000 Mons, Belgium}
\affiliation{Service de Physique Nucl\'{e}aire et Subnucl\'{e}aire,
Universit\'{e} de Mons--UMONS, Acad\'{e}mie universitaire Wallonie-Bruxelles,
Place du Parc 20, 7000 Mons, Belgium}

\date{\today}

\begin{abstract}
The Lagrange-mesh method is a powerful method to solve eigenequations written in configuration space. It is very easy to implement and very accurate. Using a Gauss quadrature rule, the method requires only the evaluation of the potential at some mesh points. The eigenfunctions are expanded in terms of regularized Lagrange functions which vanish at all mesh points except one. It is shown that this method can be adapted to solve eigenequations written in momentum space, keeping the convenience and the accuracy of the original technique. In particular, the kinetic operator is a diagonal matrix. Observables in both configuration space and momentum space can also be easily computed with a good accuracy using only eigenfunctions computed in the momentum space. The method is tested with Gaussian and Yukawa potentials, requiring respectively a small or a great mesh to reach convergence.
\end{abstract}

\maketitle

\section{Introduction}
\label{sec:intro} 

There are few three-dimensional problems in quantum mechanics which allow a complete analytical solution for any value of the orbital angular momentum (the $S$-wave channel is very similar to a simpler one-dimensional equation) \cite{flug99}. So, numerous methods have been developed to solve numerically with a high accuracy the eigenvalue equations associated with various systems. Among these techniques, the Lagrange-mesh method (LMM), which is especially easy to implement, can produce very accurate results. First created to compute eigenvalues and eigenfunctions of a two-body Schr\"{o}dinger equation \cite{BH-86,VMB93,Ba-95,hess02,Baye06,baye08}, it has been extended to treat semirelativistic Hamiltonian \cite{sema01,brau02,buis05,lag1}. The trial eigenstates are developed in a basis of particular functions, the Lagrange functions, which vanish at all mesh points, except one. Once the potential is known in the configuration space, its matrix elements are simply the potential values at the mesh points, if they are computed with an associated Gauss quadrature. At first sight, this method could look like a discrete variational method, but this is absolutely not the case since the eigenfunctions can be computed at any position. Because of the use of the Gauss quadrature scheme, the method is not variational but the results are too a large extent independent of the sole nonlinear parameter of the method fixing the physical scale of the system. Generally, a great accuracy can be reached with a small mesh \cite{baye02}. 

At the beginning, the LMM was developed in the configuration space. Recently, it has been shown that the Fourier transform of the eigenfunctions computed in the configuration space can easily be obtained with a good accuracy in the physical domain of the momentum space \cite{lacr11}. Moreover, observables in this space can easily be computed with a good accuracy using only matrix elements and eigenfunctions in the configuration space. But, for some particular problems, it can be preferable to work in the momentum space. This is the case when the potential presents discontinuities in the configuration space \cite{karr10} or when the potential is given in the momentum space. In this last case, if it is possible to use the LMM by computing first the Fourier transform of the potential, we will show here that the LMM can be adapted to solve the eigenequations directly in momentum space. Observables in both configuration and momentum spaces can also be computed with a good accuracy using only eigenfunctions computed in momentum space. Moreover, we will show that the new LMM can provide these types of data very efficiently and very easily, using again the fundamental properties of the Lagrange functions. Let us note that the method presented here relies on a mesh of points built with the zeros of a Laguerre polynomial, but a general procedure for deriving other Lagrange meshes related to orthogonal or non-orthogonal bases has also been developed \cite{baye99}. 

Several methods exist to solve eigenequations in momentum space. For instance, iterative procedures have been developed \cite{regi84,rodr88}. They are quite accurate but resort finally to numerical integrations on a mesh. Direct computations on a mesh are easier to carry out, but they require very large mesh if a good quadrature rule is not used \cite{karr10}. As we will see, the LMM in momentum space is very easy to implement and can also give accurate results. In order to fix the notations, the eigenequations in momentum space are presented in Sec.~\ref{sec:eigen}. The LMM adapted in momentum space is described in Sec.~\ref{sec:method} with some details in order that the paper is self-contained. Test calculations are presented in Sec.~\ref{sec:num} for two potentials with very different properties of convergence. Some concluding remarks are given in Sec.~\ref{sec:conclu}.

\section{Eigenequations in position and momentum spaces}
\label{sec:eigen}

Let us consider the following eigenequation ($\hbar = c = 1$)
\begin{equation}
\label{eigeneq}
\left[ T(\vec p\,^2) + V(r) \right] |\phi\rangle = E\, |\phi\rangle,
\end{equation}
in which the kinetic part and the potential depend respectively on the relative square momentum $\vec p\,^2$ and on the radial distance $r=|\vec r\,|$ between the particles. The wavefunctions in configuration space $\phi_r(\vec r\,) = \langle \vec r\, | \phi\rangle$ and in momentum space $\phi_p(\vec p\,) = \langle \vec p\, | \phi\rangle$ can be written using the spherical representation
\begin{align}
\label{phir}
\phi_r(\vec r\,)&= {\cal R}_{n l}(r)\, Y_{lm}(\hat r), \\
\label{phip}
\phi_p(\vec p\,)&= {\cal P}_{n l}(p)\, \widetilde Y_{l m}(\hat p),
\end{align}
where $\widetilde Y_{l m}(\hat x)= i^l\, Y_{l m}(\hat x)$ is a modified spherical harmonic \cite{var}, $\hat x=x/|\vec x\,|$ and $n$ is the number of nodes at finite distance. These functions are linked by the following Fourier transforms \cite{saku93}
\begin{align}
\label{FT1}
\phi_p(\vec p\,)&= \frac{1}{(2 \pi)^{3/2}}\int \phi_r(\vec r\,)\, e^{-i \vec p \cdot \vec r}\, d\vec r, \\
\label{FT2}
\phi_r(\vec r\,)&= \frac{1}{(2 \pi)^{3/2}}\int \phi_p(\vec p\,)\, e^{+i \vec p \cdot \vec r}\, d\vec p.
\end{align}
These equations lead to  \cite{lacr11}
\begin{align}
\label{PR1}
{\cal P}_{n l}(p) &= (-1)^l \sqrt{\frac{2}{\pi}} \int_0^\infty {\cal R}_{n l}(r)\, j_l(p\, r)\, r^2\, dr, \\
\label{PR2}
{\cal R}_{n l}(r) &= (-1)^l \sqrt{\frac{2}{\pi}} \int_0^\infty {\cal P}_{n l}(p)\, j_l(p\, r)\, p^2\, dp,
\end{align}
where $j_l(x)$ is a spherical Bessel function \cite{abra65}.

Written in the momentum space, (\ref{eigeneq}) takes the following form
\begin{equation}
\label{eigeneqp}
T(\vec p\,^2)\,\phi_p(\vec p\,) + \int  V_{\textrm{FT}}(\vec p-\vec p\,')\, \phi_p(\vec p\,')\, d\vec p\,' = E\, \phi_p(\vec p\,)
\end{equation}
with $V_{\textrm{FT}}(\vec p-\vec p\,')$, the Fourier transform of $V(r)$, given by
\begin{equation}
\label{VFT}
V_{\textrm{FT}}(\vec p-\vec p\,')= \frac{1}{(2 \pi)^3}\int V(r)\, e^{-i (\vec p-\vec p\,') \cdot \vec r} d\vec r.
\end{equation}
This potential is a continuous function of the momentum, even if parts of the interaction in configuration space present discontinuities. One can think of square well or Dirac delta function (repulsive only in a 3D-space). As the potential depends only on $r$, we have $V_{\textrm{FT}}(\vec p-\vec p\,')=V_{\textrm{FT}}(|\vec p-\vec p\,'|)$ and (\ref{VFT}) becomes \cite{grad80}
\begin{equation}
\label{VFTk}
V_{\textrm{FT}}(k)= \frac{1}{2\pi^2 k} \int_0^\infty V(r)\, \sin(k\, r) \, r \, dr.
\end{equation}
Using the standard decomposition of a radial function \cite{var}, the eigenvalue equation (\ref{eigeneqp}) takes the form of an integral equation for the wavefunction ${\cal P}_{n l}(p)$ 
\begin{equation}
\label{eigenP}
T(p^2)\, {\cal P}_{n l}(p) + \int_0^\infty V_l(p,p')\, {\cal P}_{n l}(p')\, {p'}^2\, dp' = E\, {\cal P}_{n l}(p)
\end{equation}
with the partial potentials
\begin{equation}
\label{Vlpq}
V_l(p,p')= 2\pi \int_{-1}^{+1} P_l(t) \, V_{\textrm{FT}}\left(\sqrt{p^2+{p'}^2-2 p p' t}\right) dt .
\end{equation}
The Legendre polynomial $P_l(t)$ depends on the variable $t=\hat p\cdot \hat p'$.

For a Schr\"odinger equation, the kinetic operator is given by
\begin{equation}
\label{TNR}
T(p^2)=\frac{p^2}{2 \mu} \quad \textrm{with} \quad \mu=\frac{m_1 m_2}{m_1 + m_2}
\end{equation}
and the eigenvalue $E$ is the binding energy of the system. In a spinless Salpeter Hamiltonian, the kinetic operator takes the following form
\begin{equation}
\label{TSR}
T(p^2)=\sqrt{p^2+m_1^2}+\sqrt{p^2+m_2^2}
\end{equation}
and the eigenvalue $E$ is the mass of the system. This kind of Hamiltonian is sometimes denoted semirelativistic since it is not a covariant formulation. This equation can be considered as a Schr\"odinger equation with its nonrelativistic kinetic part replaced by a relativistic counterpart. More rigorously, it is obtained from the covariant Bethe-Salpeter equation \cite{salp51} with the following approximations: Elimination of any dependences on timelike variables and neglect of particle spin degrees of freedom as well as negative energy solutions \cite{grei94}. The spinless Salpeter Hamiltonian is often used in hadronic physics to study bound states of quarks or gluons \cite{godf85,fulc94,buis07,math08}. 

Within this formulation, the action of the kinetic operator is just an ordinary multiplication. So,  nonrelativistic and semirelativistic systems are computed with the same manner. Moreover, more complicated kinetic parts, with momentum-dependent masses \cite{szcz96,llan00,agui11}, can be equally treated. Though the formulations in the configuration and momentum spaces are completely equivalent, this does not mean that the technical difficulties to solve the eigenequations are the same in both spaces.

If the potential is known in the configuration space, (\ref{VFTk}) and (\ref{Vlpq}) allows the computations of the partial potentials $V_l(p,p')$. $V_{\textrm{FT}}(k)$ can also be directly obtained from a physical theory, like a field theory naturally written in the momentum space. One can think of effective potentials obtained from Feynman diagrams, for instance. It is not always possible to obtain an analytical form for (\ref{Vlpq}), but such a numerical integration can be rapidly and accurately performed. 

\section{Method in momentum space}
\label{sec:method}

\subsection{Lagrange functions}
\label{ssec:Lagfunc}

The LMM relies on the existence of a $N$-point mesh $\{x_i\}$, which is associated with an orthonormal set of $N$ indefinitely derivable functions $\bar f_j (x)$, called the Lagrange function \cite{BH-86,VMB93,Ba-95}. Each function $\bar f_j(x)$ satisfies the Lagrange conditions, 
\begin{equation}
\label{fLM}
\bar f_j (x_i) = \lambda_i^{-1/2} \delta_{ij},
\end{equation}
that is to say it vanishes at all mesh points except one. The $x_i$ and $\lambda_i$ are respectively the abscissae and the weights of a Gauss quadrature formula
\begin{equation}
\label{Gaussquad} 
\int^{\infty}_0 g(x)\, dx \approx \sum_{k=1}^{N} \lambda_k\, g(x_k).
\end{equation}
Several quadratures are possible, but, as we work with wavefunctions depending only on the module of a variable, we consider the case of the Gauss-Laguerre quadrature whose domain of interest is $[0,\infty[$.
The Gauss formula (\ref{Gaussquad}) is exact when $g(x)$ is a polynomial of degree $2N-1$ at most, multiplied by $\exp (-x)$. The mesh points $x_i$ are the zeros of a Laguerre polynomial of degree $N$: $L_N (x_i)=0$ \cite{BH-86}. These zeros can be determined with a high precision with usual methods to find the roots of a polynomial \cite{numrec} (the \emph{Mathematica} expression \texttt{Root} does efficiently the job) or as the eigenvalues of a particular tridiagonal matrix \cite{golu69}. The weights can be computed by the following formula \cite{baye02}
\begin{equation}
\label{lambi}
\ln \lambda_i = x_i - \ln x_i + 2 \ln\Gamma(N+1)-\sum_{j\ne i=1}^N \ln(x_i-x_j)^2.
\end{equation}

Wavefunctions $p\, {\cal P}_{n l}(p)$ vanish at the origin. As it is not the case for the original Lagrange functions, it is preferable to use the regularized Lagrange functions whose explicit form is given by $f_i (x) = (x/x_i)\bar f_i (x)$, that is to say \cite{VMB93,Baye06}
\begin{equation}
\label{fLMreg}
f_i (x) = (-1)^i x_i^{-1/2} x (x-x_i)^{-1} L_N (x) \exp (-x/2).
\end{equation}
Such a function satisfies (\ref{fLM}) and $f_i (0) = 0$ (see Eq.~\ref{Pexp2}).

\subsection{Matrix equation}
\label{sec:matrix}

The use of the LMM in configuration space is described in \cite{BH-86,VMB93,Ba-95,Baye06,baye08,hess02,sema01,lacr11}. We will present here the formulation for the integral equation (\ref{eigenP}) within the LMM. The idea is to expand the wavefunction ${\cal P}_{n l}(p)$ with the regularized Lagrange functions in such a way that a trial state $|\psi\rangle$ is written
\begin{equation}
\label{Pexp}
|\psi\rangle = \sum_{j=1}^N C_j\, |f_j \rangle \quad \textrm{with} \quad \langle \vec p\, |f_j \rangle =
\frac{f_j (p/h)}{\sqrt{h}\,p}\, \widetilde Y_{l m}(\hat p).
\end{equation}
This formula is identical to formula (6) in \cite{lacr11} with just the replacement of the variable $r$ by $p$. The coefficients $C_j$ are linear variational parameters and the scale factor $h$ is a nonlinear parameter aimed at adjusting the mesh to the domain of physical interest. The parameter $h$ has here the dimension of a momentum ($h$ plays the same role in \cite{lacr11} but has the dimension of a distance). Contrary to some other mesh methods, the wavefunction is also defined between mesh points by (\ref{fLMreg}) and (\ref{Pexp}). For a good value of $h$ and a sufficiently high value of $N$ (see sec.~\ref{sec:num}), the function
\begin{equation}
\label{Pexp2}
\bar {\cal P}_{n l}(p)= \sum_{j=1}^N C_j\, \frac{f_j (p/h)}{\sqrt{h}\,p}
\end{equation}
can be a good approximation of the exact function ${\cal P}_{n l}(p)$. Note that an advantage of the LMM is that the value for $h$ has not to be known with a great accuracy. It is sufficient that this value be located within a given interval, as we will check in the following. 

Basis states $|f_i\rangle$ built with the regularized Lagrange functions are not exactly orthogonal. But, at the Gauss approximation, we have $\langle f_j|f_i\rangle = \delta_{ji}$. So, in the following, all mean values will be performed using the Gauss quadrature formula (\ref{Gaussquad}). Inserting expansion (\ref{Pexp2}) in (\ref{eigenP}) gives 
\begin{equation}
\label{Pexp3}
T(h^2 x^2) \sum_{j=1}^N C_j\, \frac{f_j(x)}{x} + \sum_{j=1}^N C_j\, h^3\, \sqrt{\lambda_j}\, x_j\, V_l(h\, x_j, h\, x) = E \sum_{j=1}^N C_j\, \frac{f_j(x)}{x},
\end{equation}
where $x=p/h$ is a dimensionless variable. We can now multiply this equation by $x\, f_j(x)$ and integrate on $[0,\infty[$ with again the Gauss quadrature formula (\ref{Gaussquad}). Finally, we obtain
\begin{equation}
\label{Pexp4}
\sum_{j=1}^N C_j \left[ T(h^2 x_i^2)\,\delta_{ij} + h^3\, \sqrt{\lambda_i\,\lambda_j}\, x_i\,x_j\, V_l(h\, x_i, h\, x_j) - E\, \delta_{ij} \right] = 0.
\end{equation}
The Hamiltonian matrix is symmetric since $V_l(p,p')=V_l(p',p)$. A similar expression is obtained for calculations with the LMM in the configuration space for a nonlocal potential \cite{hess02}. With the LMM in momentum space, the solution of a quantum equation reduces (as it is often the case) to the determination of eigensolutions of a given matrix. So, once the partial potentials are known, (\ref{Pexp4}) shows that this method is very easy to implement. The computations of the Fourier transform of the wavefunction and of observables is no more complicated, as presented in the following. 

In Sec.~\ref{sec:num}, we will study the convergence of the method as a function of the scale parameter $h$ and the number of mesh points $N$ for nonrelativistic an semirelativistic kinematics. Let us note that an automatic determination of $h$ has been developed for the LLM in the configuration space \cite{brau98,lacr11}. But, the generalization of such a technique to the LLM in the momentum space is very difficult due to the nonlocal nature of the interaction in (\ref{eigenP}). We will cross-check our results by comparing eigenvalues and mean values of observables computed with the LMM in both configuration and momentum spaces. Moreover, a wavefunction computed in the momentum space will be compared with the Fourier transform of the corresponding wavefunction computed in the configuration space, and vice versa.

\subsection{Fourier transform}
\label{sec:FT}

It has been shown in \cite{lacr11} that a good approximation of a function ${\cal P}_{n l}(p)$ can be obtained from the Fourier transform of the corresponding solution computed in the configuration space by the LLM. This can be performed by taking benefit of the very special properties of the regularized Lagrange functions. Similarly, a good approximation of the function ${\cal R}_{n l}(r)$ can be obtained from the Fourier transform of a solution computed in the momentum space by the LLM. Using the Gauss quadrature rule (\ref{Gaussquad}) with the Lagrange condition (\ref{fLM}), the integral (\ref{PR2}) for a given trial state (\ref{Pexp2}) simply reduces to
\begin{equation}
\label{FT7}
\bar {\cal R}_{n l}(r) = (-1)^l \sqrt{\frac{2}{\pi}} \, h^{3/2} \sum_{i=1}^N C_i\, \sqrt{\lambda_i}\, 
x_i\, j_l(h\, x_i\, r).
\end{equation}
This formula is identical to formula (22) in \cite{lacr11} with just the replacement of the variable $r$ by $p$. This results from the similarity of (\ref{FT1}) and (\ref{FT2}), and the choice of expansion (\ref{Pexp}). For a sufficiently high value of $N$ (see sec.~\ref{sec:num}), $\bar {\cal R}_{n l}(r)$ can be a very good approximation of the genuine function ${\cal R}_{n l}(r)$ in the configuration space. Above a critical value of $r$, generally in the asymptotic tail of  $\bar {\cal R}_{n l}(r)$, this function can present large unphysical rapid oscillations. These oscillations do not develop in $\bar {\cal P}_{n l}(p)$, because they are damped by the rapid decrease of the regularized Lagrange functions.

\subsection{Mean values of momentum-dependent observables}
\label{sec:meanp}

The mean value of the operator $U(p)$ for a trial state $|\psi \rangle$ is given by
\begin{equation}
\label{meanU1}
\langle\psi |U(p)|\psi \rangle = \sum_{i,j=1}^N C_i\, C_j\,\langle f_i |U(p)|f_j \rangle.
\end{equation}
Using the Lagrange condition (\ref{fLM}) and the Gauss quadrature (\ref{Gaussquad}), this integral reduces to
\begin{equation}
\label{meanU2}
\langle\psi |U(p)|\psi \rangle = \sum_{j=1}^N C_j^2\, U(h\, x_j).
\end{equation}
If $U$ is the identity, we recover the normalization condition as expected. As we will see in sec.~\ref{sec:num}, a very good accuracy can be reached for the mean values $\langle U(p) \rangle$.

\subsection{Mean values of radial observables}
\label{sec:meanr}

The mean value of the operator $K(r)$ for a trial state $|\psi \rangle$ is given by
\begin{equation}
\label{meanK2}
\langle\psi |K(r)|\psi \rangle = \sum_{i,j=1}^N C_i\, C_j\, \langle f_i |K(r)|f_j \rangle.
\end{equation}
The method to compute matrix elements $\langle f_i |K(r)|f_j \rangle$ relies on the fact that $\vec r\,^2=-\vec \nabla^2_{\vec p}$ in the momentum space \cite{luch90}. Let us first look at the matrix $P$ whose elements are $P_{ij}=\langle f_i|\vec r\,^2|f_j\rangle$.  With (\ref{Gaussquad}), these matrix elements are given by
\begin{equation}
\label{r2}
P_{ij}= \frac{1}{h^2} \left( t_{ij} + \frac{l(l+1)}{x_i^2}\delta_{ij}\right),
\end{equation}
where
\begin{equation}
\label{tij}
t_{ij} = \int^{\infty}_0 f_i (x) \left( -\frac{d^2}{dx^2} \right)
f_j (x)\ dx \approx - \lambda_i^{1/2} f_j'' (x_i).
\end{equation}
This expression is exact for some Lagrange meshes, but this is not the case for the regularized Laguerre mesh. An exact expression can easily be obtained (see the Appendix in \cite{VMB93}). However, as shown in \cite{Ba-95}, it is preferable to use the approximation (\ref{r2})-(\ref{tij}). The quantities $t_{ij}$ are then easy to obtain and read \cite{Ba-95}
\begin{equation}
\label{tijbis}
t_{ij} = \left \{ \begin{array}{ll}
(-)^{i-j} (x_i x_j)^{-1/2} (x_i+x_j)(x_i-x_j)^{-2}
& (i \neq j), \\
(12x_i^2)^{-1} [4 + (4N + 2) x_i - x_i^2]
& (i = j). \end{array} \right.
\end{equation}
If $P^D$ is the diagonal matrix formed by the eigenvalues of $P$, we have \begin{equation}
\label{step2}
P = S\,P^D\,S^{-1},
\end{equation}
where $S$ is the transformation matrix composed of the normalized eigenvectors. Let us call $K^D$ the diagonal matrix obtained by taking the function $K(\sqrt{x})$ of all diagonal elements of $P^D$ (remember that $P$ is linked to the matrix elements of $r^2$, not $r$). The numbers $\langle f_i |K(r)|f_j \rangle$ are well approximated by the elements of the matrix $K$ obtained by using the transformation (\ref{step2}): $K = S\,K^D\,S^{-1}$. As we will see in sec.~\ref{sec:num}, a very good accuracy can be reached for the mean values $\langle K(r) \rangle$.

\section{Numerical results}
\label{sec:num}

\subsection{Gaussian potential}
\label{sec:gauss}

We first consider a Gaussian potential whose Fourier transform is well known  \cite{grad80}
\begin{equation}
\label{Vkgauss}
V(r) = -a\, \exp\left(- b^2\, r^2\right) \Leftrightarrow V_{\textrm{FT}}(k)= -\frac{a}{8\,\pi^{3/2} b^3 } \exp\left(-\frac{k^2}{4\,b^2}\right).
\end{equation}
Let us note that the asymptotic behavior of the potential is similar in the configuration and momentum spaces. Using (\ref{Vlpq}), we can compute 
\begin{equation}
\label{VlGauss}
V_l(p,p')= -\frac{a}{2^{l+2}\sqrt{\pi}\,b^3} \exp\left( -\frac{p^2+{p'}^2}{4\,b^2}\right) \sum_{k=0}^{[l/2]}
(-1)^k \binom{l}{k} \binom{2 l-2 k}{l} \left[ E_{2 k-l}\left( -\frac{p\, p'}{2\,b^2} \right) +
(-1)^{l-2 k}  E_{2 k-l}\left( \frac{p\, p'}{2\,b^2} \right) \right],
\end{equation}
where $E_{m}(z)$ is an exponential integral \cite{abra65} and $[y]$ means the integer part of $y$. This type of potentials is the prototype for a short-range interaction and can be used in many field of physics. 

With a nonrelativistic kinematics, we can write
\begin{equation}
\label{Egauss}
E = \frac{b^2}{2 \mu} \epsilon,
\end{equation}
where $\epsilon$ is the solution of the dimensionless Hamiltonian
\begin{equation}
\label{Hdimgauss}
\tilde H = \vec q\,^2 - g\, e^{-x^2} \quad \textrm{with} \quad g = \frac{2\,\mu\, a}{b^2}.
\end{equation}
We choose to study the performance of the LMM in momentum space with the value $g=15$. In this case, two bound states exist: $(n,l)=(0,0)$ and $(0,1)$. 

\begin{figure}[htb]
\includegraphics*[width=0.45\textwidth]{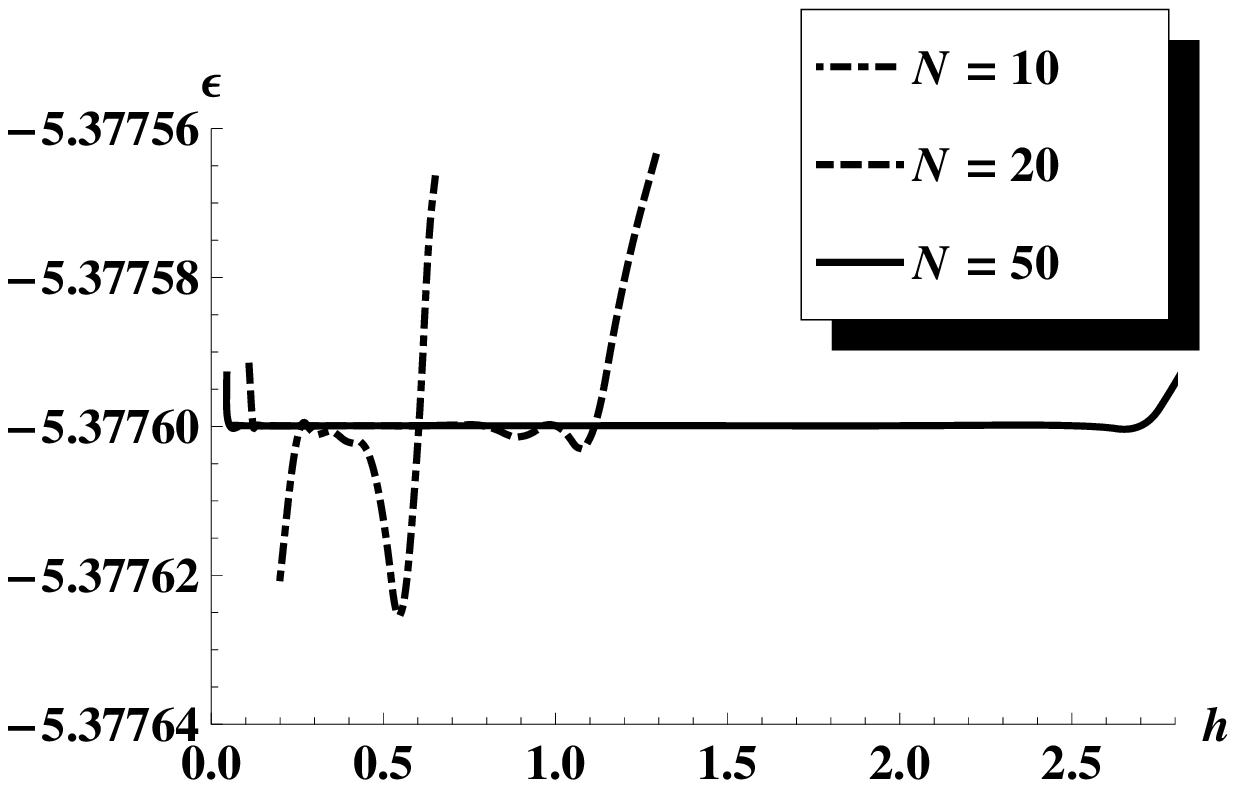}
\includegraphics*[width=0.45\textwidth]{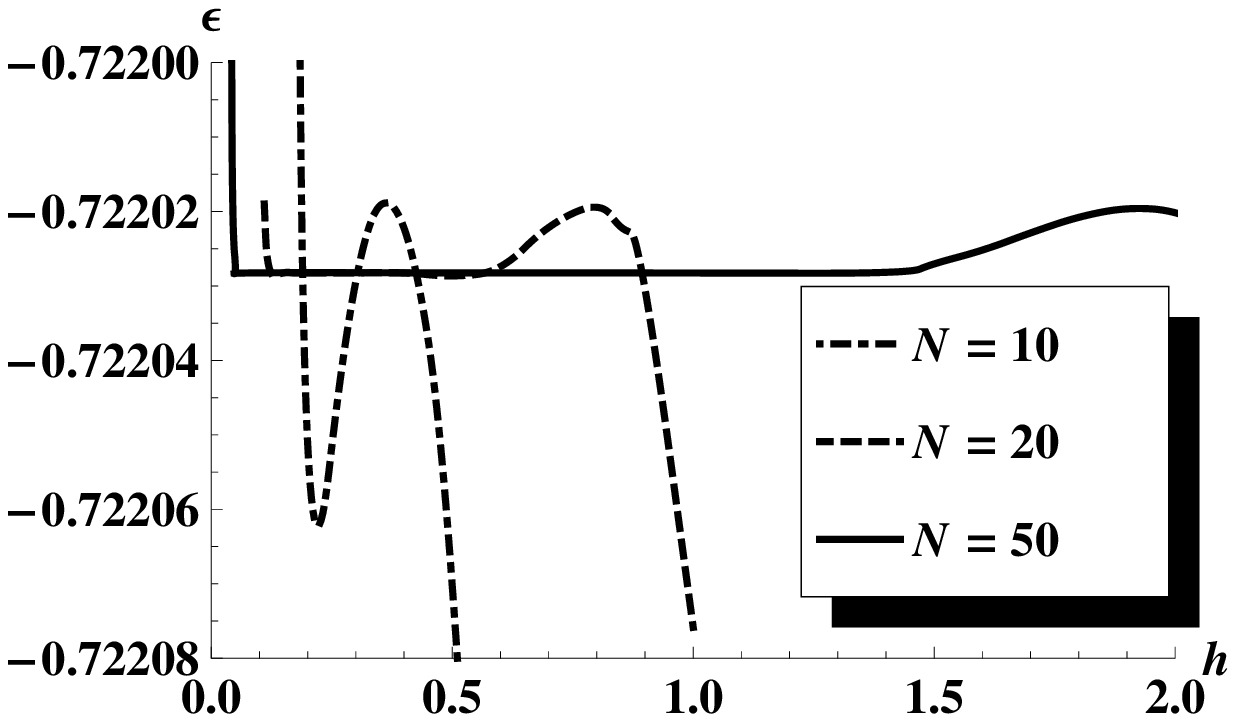}
\caption{Eigenvalues $\epsilon$ of (\ref{Pexp4}) for the dimensionless Hamiltonian (\ref{Hdimgauss}) with $g=15$, as a function of $h$ for three values of $N$. Left: Ground state. Right: Excited state with $(n,l)=(0,1)$. The scale for ordinates is the same for both graphics.
\label{fig:NRgauss}}
\end{figure}

\begin{figure}[htb]
\includegraphics*[width=0.45\textwidth]{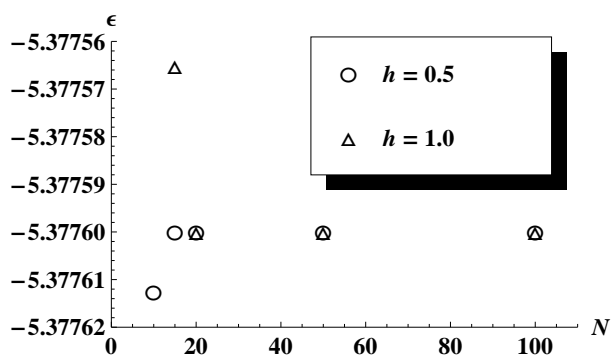}
\caption{Ground state eigenvalues $\epsilon$ of (\ref{Pexp4}) for the dimensionless Hamiltonian (\ref{Hdimgauss}) with $g=15$, as a function of $N$ for two values of $h$. For $N=10$ and $h=1.0$, the value of $\epsilon$ is $-5.37859$ far below the range of the graph. 
\label{fig:NRgauss2}}
\end{figure}

Let us look at the results obtained for $\tilde H$ (\ref{Hdimgauss}) by solving (\ref{Pexp4}). The variation of the eigenvalue $\epsilon$ as a function of the nonlinear parameter $h$ is presented for the two states in Fig.~\ref{fig:NRgauss}. For a value of $N$ as small as 20, a plateau is present with abrupt variations of the eigenvalue at the borders. The length of the plateau increases with the number $N$ of mesh points. For the excited states, the convergence is usually slower than for the ground states: The plateau is less extended and the variations around the plateau are larger. Even with $N=10$, very good results can be obtained provided $h$ is taken in the small corresponding quasi-plateau (the fluctuation in the rapid variation). Note that the flatness of the plateau is a criterion obviously depending on the accuracy one wants to reach. 
It is shown on Fig.~\ref{fig:NRgauss2} that eigenvalues, corresponding to different values of $h$ taken in plateaus, converge towards the same value as $N$ increases. This convergence can be achieved from above or from below. All these results clearly show the non variational nature of the LMM. 

It is remarkable that, even for small values of $N$, the value of $h$ must not be determined with a high precision. This is a great advantage for the LMM. Nevertheless, if $h$ is too small, a significant part of the wavefunction is not covered by the points of the mesh. When $h$ is too large, all points of the mesh are located in the asymptotic tail of the wavefunctions and it is then impossible to obtain good eigenvalues. Let us note that the same behaviors are observed for the LMM in configuration space \cite{BH-86,Ba-95,sema01}. 

Eigenvalues and some observables have been calculated with the LMM in both configuration and momentum spaces. Results are presented and compared in Table~\ref{tab:GaussNRcomp} for the ground state only, but results are similar for the excited state. A very good accuracy can be obtained for the computation in momentum space, even for small values of $N$. For most of the physical problems, the required precision can probably be reached with $N=20$. It seems that position-dependent observables converge more slowly than momentum-dependent observables. We have checked that the situation is the opposite for LMM calculations in configuration space.

In principle, we must have $\langle \tilde H \rangle = \epsilon$. This is not the case since momentum-dependent and radial observables are not computed with the same method (see Sec.~\ref{sec:meanp} and \ref{sec:meanr}). We can see in Table~\ref{tab:GaussNRcomp} that the difference between the two quantities is around the accuracy of the eigenvalues $\epsilon$.

\begin{center}
\begin{table}[htb]
\caption{Eigenvalues $\epsilon$ and mean values of some observables for the ground state of $\tilde H=\vec q\,^2 + U(x)$ with $U(x)=-g\, e^{-x^2}$ (\ref{Hdimgauss}) and $g=15$, as a function of $N$ for $h=0.5$. Computations in momentum (Mom.) space are compared with accurate results obtained in configuration (Conf.) space for $N=100$ and $h=0.4$. In order to check the mean values, $\langle \tilde H \rangle$ (which must be equal to $\epsilon$) is computed by $\langle \vec q\,^2 \rangle + \langle U(x) \rangle$.
\label{tab:GaussNRcomp}}
\begin{tabular}{ccccc}
\hline\hline
 & Conf. & \multicolumn{3}{c}{Mom.} \\
 &  & $N=10$ & $N=20$ & $N=50$ \\
\hline
$\epsilon$ & $-5.3775999070684$ & $-5.3776125307238$ & $-5.3775999078195$ & $-5.3775999070682$ \\
$\langle \vec q\,^2 \rangle$ & $3.74063887622353$ & $3.74063826403371$ & $3.74063885577063$ & $3.74063887622358$  \\
$\langle \vec q\,^4 \rangle$ & $26.50642515647$ & $26.50643641212$ & $26.50642516641$ & $26.50642515646$  \\
$\langle x \rangle$ & $0.7134620$ & $0.7135030$ & $0.7134650$ & $0.7134620$  \\
$\langle U(x) \rangle$ & $-9.1182387832920$ & $-9.1182424774223$ & $-9.1182387633200$ & $-9.1182387832920$  \\
$\langle \tilde H \rangle$ & $-5.3775999070685$ & $-5.3776042133885$ & $-5.3775999075493$ & $-5.3775999070684$  \\
\hline\hline
\end{tabular}\\
\end{table}
\end{center}

\begin{figure}[htb]
\includegraphics*[width=0.45\textwidth]{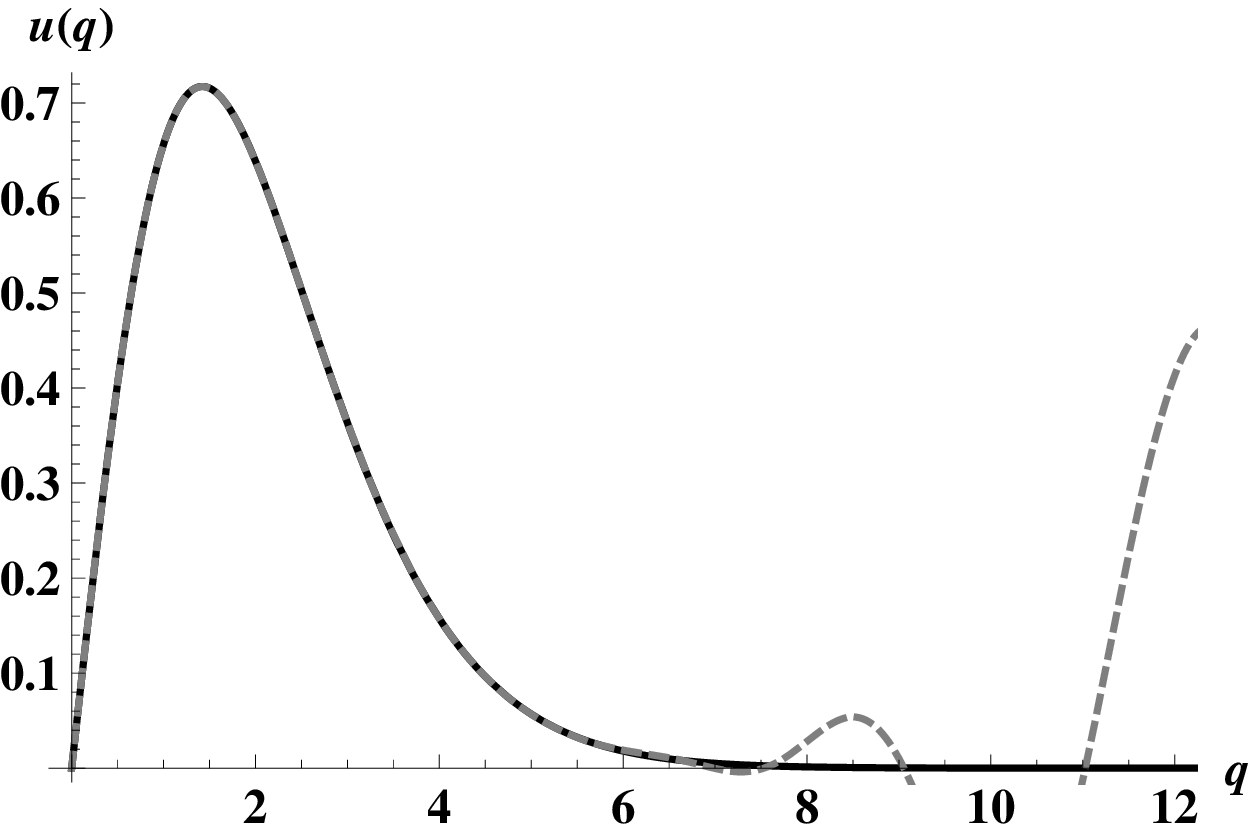}
\includegraphics*[width=0.45\textwidth]{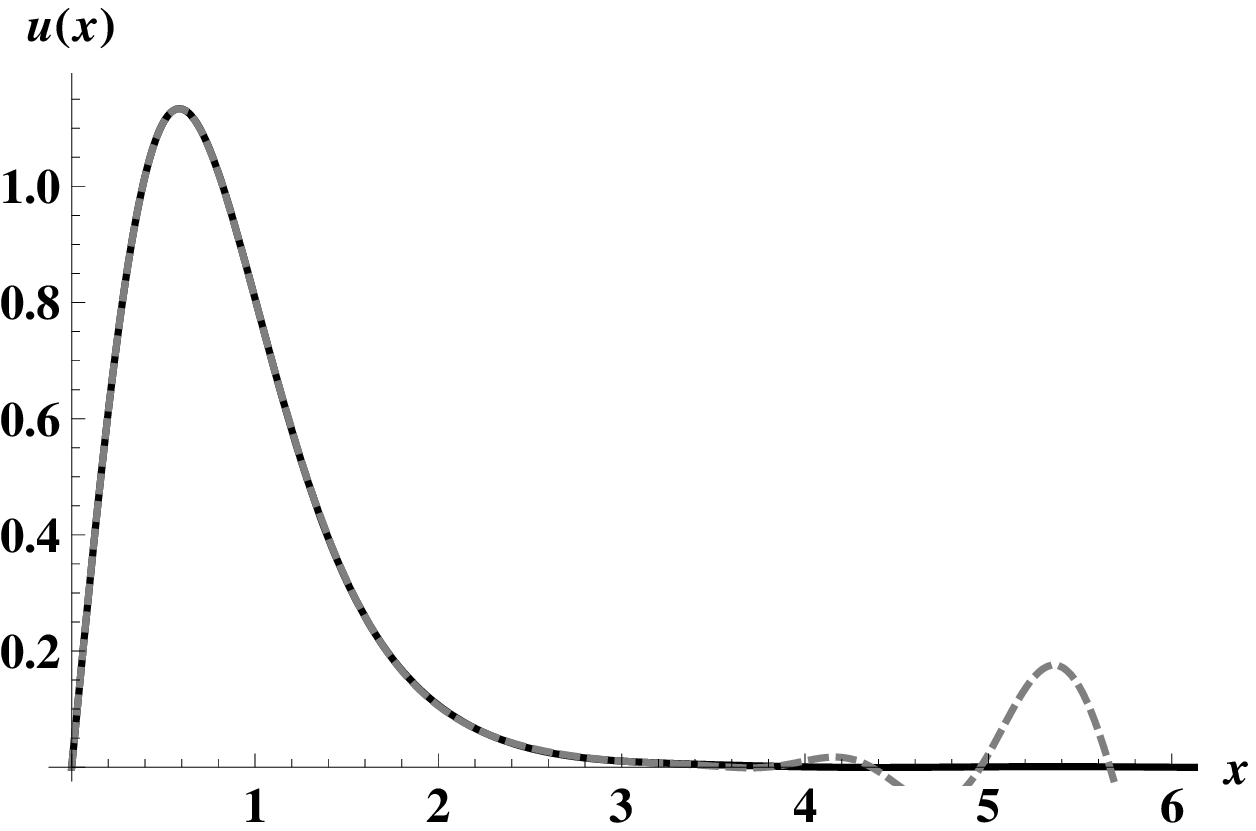}
\caption{Ground state of the dimensionless Hamiltonian (\ref{Hdimgauss}) with $g=15$ (only positive values are shown). Left: $u(q)=q\,{\cal P}_{0 0}(q)$ [black solid line: wavefunction obtained directly in the momentum space -- gray dashed line: wavefunction computed by Fourier transform of the wavefunction obtained directly in the configuration space]. Right: $u(x)=x\,{\cal R}_{0 0}(x)$ [black solid line: wavefunction obtained directly in the configuration space -- gray dashed line: wavefunction computed by Fourier transform of the wavefunction obtained directly in the momentum space]. $h=0.5$ and $N=20$ ($h=0.4$ and $N=20$) for the computation in momentum (configuration) space.
\label{fig:gausswf}}
\end{figure}

The wavefunctions produced by the LMM in both spaces have also been compared. A typical example is shown on Fig.~\ref{fig:gausswf}. On the left part, the wavefunction obtained directly in the momentum space and the wavefunction computed by the Fourier transform of the wavefunction obtained directly in the configuration space are superposed. The situation is the opposite on the right part. In both cases, the agreement is very good for low values of the arguments, but the Fourier transform wavefunctions present large unphysical oscillation for larger values of the argument. The starting points of these oscillations can be rejected to high values of the argument in the asymptotic tail, but it is then necessary to work with greater values of $N$, typically $N \gtrsim 100$ \cite{lacr11}. A good representation of the wavefunction in momentum space can be obtained with a small mesh only by working directly in this space. This an advantage of the LMM in momentum space.

\begin{figure}[htb]
\includegraphics*[width=0.45\textwidth]{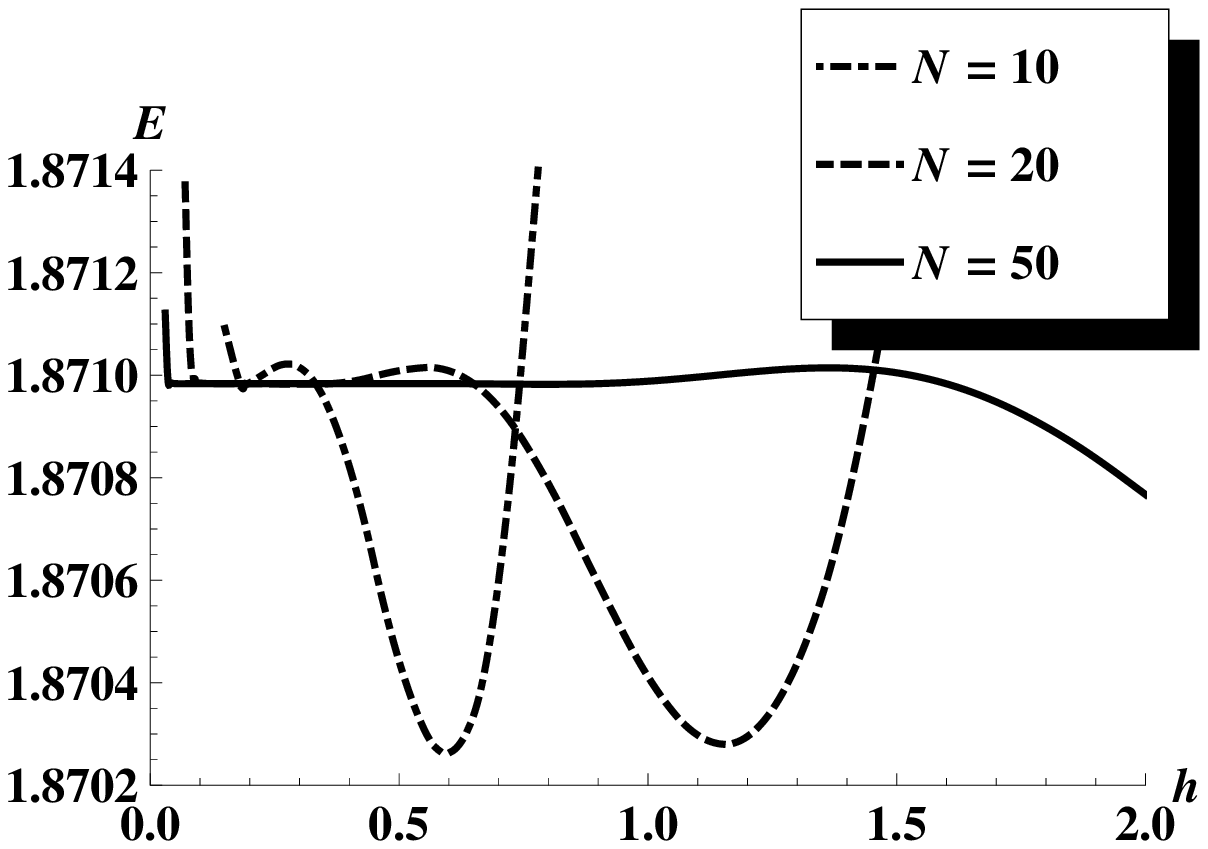}
\includegraphics*[width=0.45\textwidth]{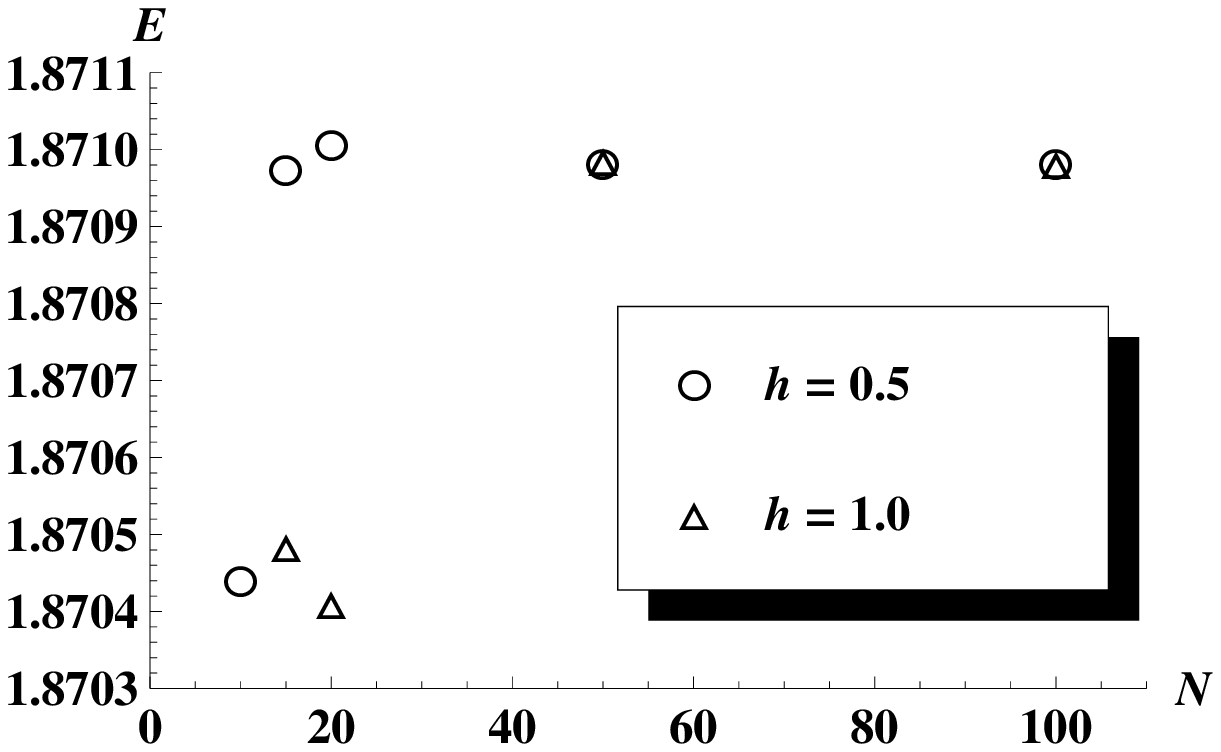}
\caption{Ground state eigenvalue $E$ of (\ref{Pexp4}) for the spinless Salpeter Hamiltonian with the Gaussian interaction considered (\ref{paramg}), as a function of $h$ and $N$. For $N=10$ and $h=1.0$, the value of $E$ is $1.8750$ far above the range of the graph.
\label{fig:srgauss}}
\end{figure}

As in the nonrelativistic case, we will not try to study a ``realistic" relativistic system. We will consider quite arbitrary, but convenient, values for the parameters,
\begin{equation}
\label{paramg}
m_1=m_2=m=1,\quad a=3, \quad b=1,
\end{equation}
for which only one bound state exists ($0 < E < 2 m$). As we can see by examining Fig.~\ref{fig:srgauss}, the behaviors of the eigenvalues as a function of $N$ and $h$ is similar for nonrelativistic and semirelativistic Hamiltonians. But, the convergence is less good for semirelativistic systems, with shorter plateaus and larger variations around the plateaus. This situation is similar in configuration space \cite{sema01,lacr11}. Various observables have been computed and are presented in Table~\ref{tab:GaussSRcomp}. A reasonable accuracy can be obtained even with a small mesh. 

\begin{center}
\begin{table}[htb]
\caption{Eigenvalues $E$ and mean values of some observables for the ground state of the spinless Salpeter Hamiltonian with the Gaussian interaction $U(r)=-a \,\exp\left(- b^2 r^2  \right)$ considered (\ref{paramg}), as a function of $N$ for $h=0.4$. Computations in the momentum (Mom.) space are compared with accurate results obtained in the configuration (Conf.) space for $N=100$ and $h=0.4$. In order to check the mean values, $\langle H \rangle$ (which must be equal to $E$) is computed by $2\langle \sqrt{\vec p\,^2+m^2} \rangle + \langle U(r) \rangle$.
\label{tab:GaussSRcomp}}
\begin{tabular}{cccccc}
\hline\hline
 & Conf. & \multicolumn{3}{c}{Mom.} \\
 &  & $N=10$ & $N=20$ & $N=50$ \\
\hline
$E$ & $1.87098362$ & $1.87044199$ & $1.87100878$ & $1.87098367$ \\
$\langle \sqrt{\vec p\,^2+m^2} \rangle$ & $1.3553804$ & $1.3542724$ & $1.3554650$ & $1.3553807$  \\
$\langle \sqrt{\vec p\,^4} \rangle$ & $3.991567$ & $3.981098$ & $3.992369$ & $3.991570$  \\
$\langle r \rangle$ & $1.73375$ & $1.71171$ & $1.73551$ & $1.73376$  \\
$\langle U(r) \rangle$ & $-0.8397772$ & $-0.8381094$ & $-0.8399212$ & $-0.8397777$  \\
$\langle H \rangle$ & $1.87098362$ & $1.87043532$ & $1.87100880$ & $1.87098367$  \\
\hline\hline
\end{tabular}\\
\end{table}
\end{center}

\subsection{Yukawa potential}
\label{sec:yuk}

Some numerical tests are also performed using the Yukawa potential whose asymptotic behavior is  very different in the configuration and momentum spaces \cite{grad80}
\begin{equation}
\label{VkYuk}
V(r) = -\frac{a}{r} \exp\left( - b\, r\right) \Leftrightarrow V_{\textrm{FT}}(k)= -\frac{1}{2\pi^2} \frac{a}{b^2+k^2}.
\end{equation}
 Using (\ref{Vlpq}), we can compute 
\begin{equation}
\label{VlYuk2}
V_l(p,p')= -\frac{a}{\pi p p'} Q_l\left( \frac{b^2+p^2+{p'}^2}{2 p p'}\right),
\end{equation}
where $Q_l(x)$ is a Legendre function of the second kind \cite{magn66}. This type of potentials is used in many field of physics, even in hadronic physics. If the interaction in a quark-antiquark system or between two gluons in the vacuum can be simulated by a funnel potential (Coulomb+linear), the confinement vanishes inside a quark-gluon plasma above a critical temperature and the interaction could turn into a Yukawa potential \cite{brau07}. 

With a nonrelativistic kinematics, we can write
\begin{equation}
\label{Eyuk}
E = \frac{b^2}{2 \mu} \epsilon,
\end{equation}
where $\epsilon$ is the solution of the dimensionless Hamiltonian
\begin{equation}
\label{Hdim}
\tilde H = \vec q\,^2 - g\, \frac{e^{-x}}{x} \quad \textrm{with} \quad g = \frac{2\,\mu\, a}{b}.
\end{equation}
We choose to study the performance of the LMM in momentum space with the value $g=10$. In this case, three bound states exist. A phenomenological approximate formula \cite{gree82} gives $E_{(n=0,l=0)}=-16.32$, $E_{(1,0)}=-0.65$ and $E_{(0,1)}=-0.22$.

\begin{figure}[htb]
\includegraphics*[width=0.45\textwidth]{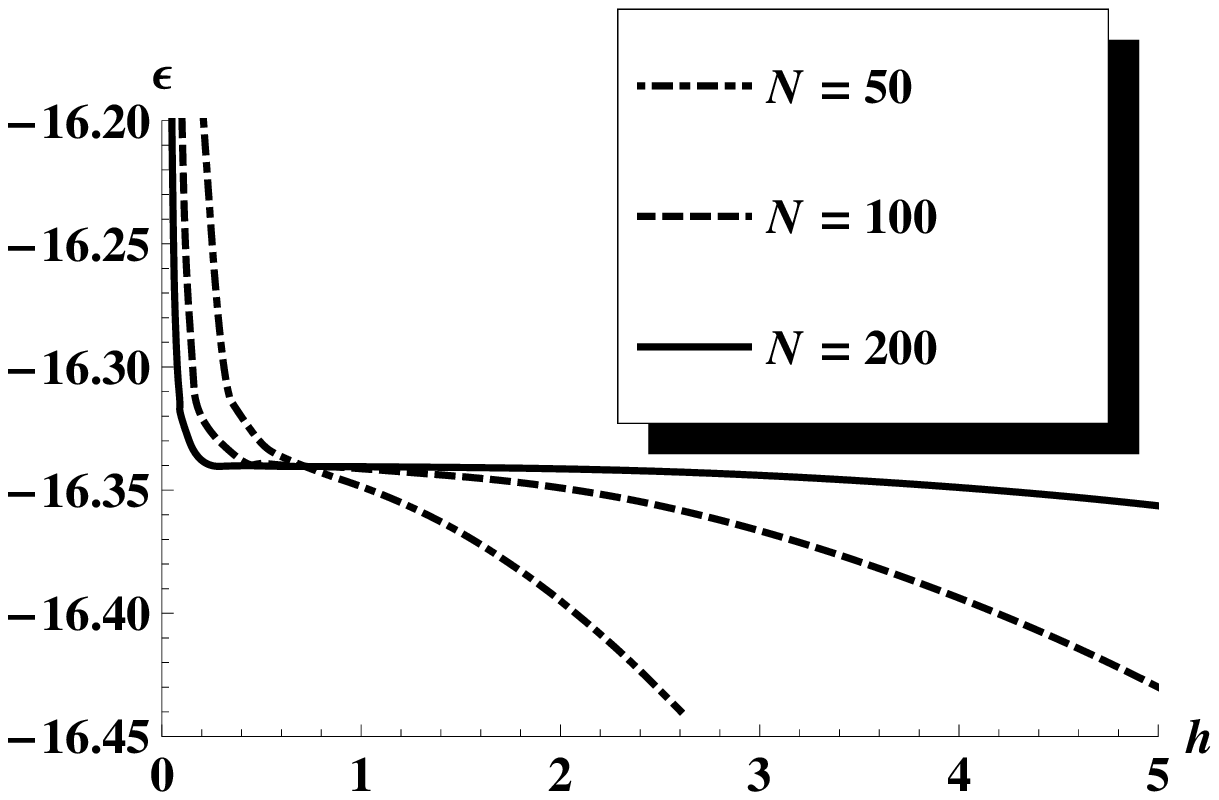}
\includegraphics*[width=0.45\textwidth]{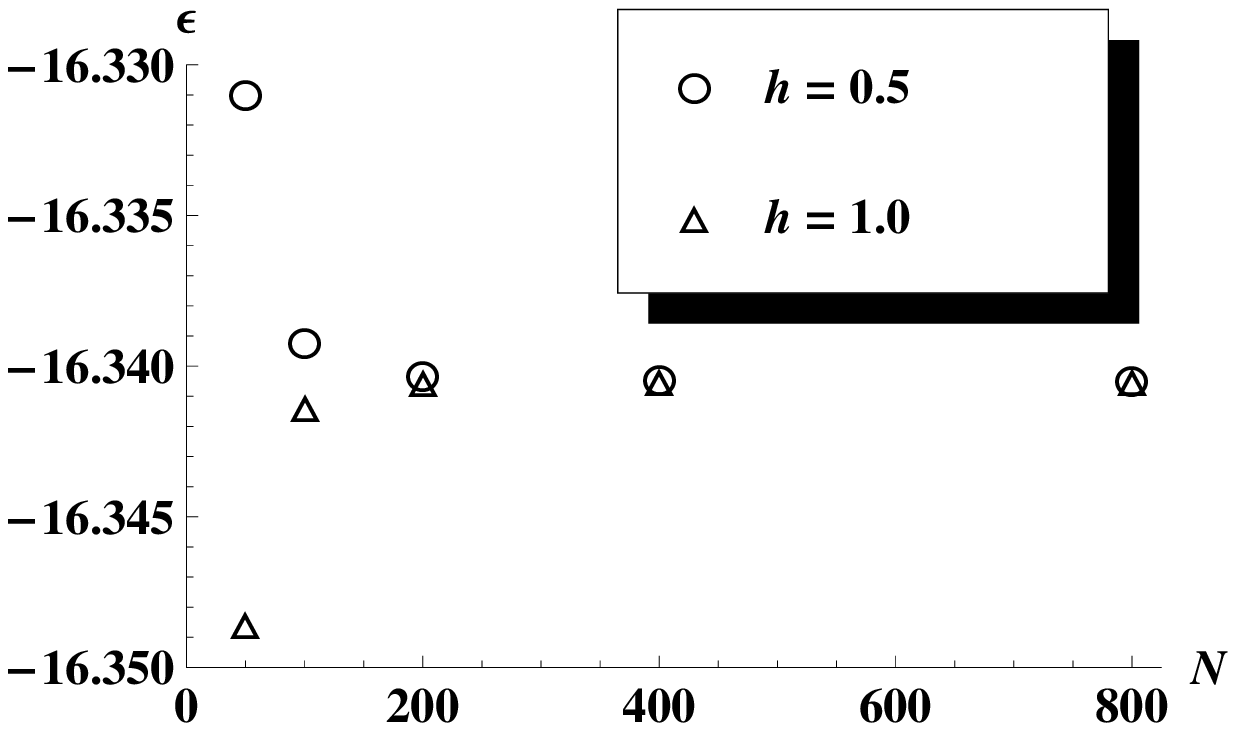}
\caption{Ground state eigenvalue $\epsilon$ of (\ref{Pexp4}) for the dimensionless Hamiltonian (\ref{Hdim}) with $g=10$. 
\label{fig:NRYuk}}
\end{figure}

On Fig.~\ref{fig:NRYuk}, we can see the behaviors of the ground state eigenvalue $\epsilon$ of $\tilde H$ (\ref{Hdim}), obtained by solving (\ref{Pexp4}), as a function of $N$ and $h$. The situation is at first sight similar to the case of the Gaussian potential, but  the convergence is much slower for the Yukawa potential. For a mesh of 50 points, a plateau does not appear clearly, just a slowing down of the variation of $\epsilon$. For computations in the configuration space, we have checked that a small plateau is already present for $N=20$. Consequently, eigenvalues computed in momentum space for two different values of $h$ will only coincide for large values of $N$. About ten times more points are necessary for computations in momentum space to reach an accuracy similar to the one of computations in configuration space. It is illustrated in Table~\ref{tab:YukNRcomp} where eigenvalues and some observables are calculated with the LMM in both configuration and momentum spaces. Again, the mean value $\langle \tilde H \rangle$, which must be equal to $\epsilon$, is computed. The disagreement for the results in momentum space shows that the convergence is not so good than for the results in configuration space. We think that this peculiarity is linked to the asymptotic behavior of the Yukawa potential in the momentum space: Its decrease like a power-law (\ref{VkYuk}) is quite slow. The wavefunction in this space is then very extended and requires its computation at high values of $p$ to be well described on its physical domain. 

\begin{center}
\begin{table}[htb]
\caption{Eigenvalues $\epsilon$ and mean values of some observables for $\tilde H=\vec q\,^2 + U(x)$ with $U(x)=-g\, e^{-x}/x$ (\ref{Hdim}) and $g=10$. Computations in configuration (Conf.) and momentum (Mom.) spaces are performed with $N=200$. The values of the parameter $h$ are arbitrarily taken into a corresponding plateau. In order to check the mean values, $\langle \tilde H \rangle$ (which must be equal to $\epsilon$) is computed by $\langle \vec q\,^2 \rangle + \langle U(x) \rangle$.
\label{tab:YukNRcomp}}
\begin{tabular}{ccccccc}
\hline\hline
 & \multicolumn{2}{c}{$n=0,\ l=0$} & \multicolumn{2}{c}{$n=1,\ l=0$} & \multicolumn{2}{c}{$n=0,\ l=1$} \\
 & Conf. & Mom. & Conf. & Mom. & Conf. & Mom. \\
 &($h=0.02$) & ($h=0.8$) & ($h=0.05$) & ($h=1.0$) &($h=0.05$) & ($h=0.5$)\\
\hline
$\epsilon$ & $-$16.340426 & $-$16.340415 & $-$0.6053933 & $-$0.6053975 & $-$0.205082327 & $-$0.205082331 \\
$\langle \vec q\,^2 \rangle$ & 23.788977 & 23.788942 & 2.95238 & 2.95241 & 2.70792857 & 2.70792862 \\
$\langle U(x) \rangle$ & $-$40.1294 & $-$40.1200 & $-$3.55778 & $-$3.55743 & $-$2.913010896 & $-$2.913010877 \\
$\langle \tilde H \rangle$ & $-$16.340426 & $-$16.331047 & $-$0.6053933 & $-$0.6050217 & $-$0.205082327 & $-$0.205082257 \\
\hline\hline
\end{tabular}\\
\end{table}
\end{center}

\begin{figure}[htb]
\includegraphics*[width=0.45\textwidth]{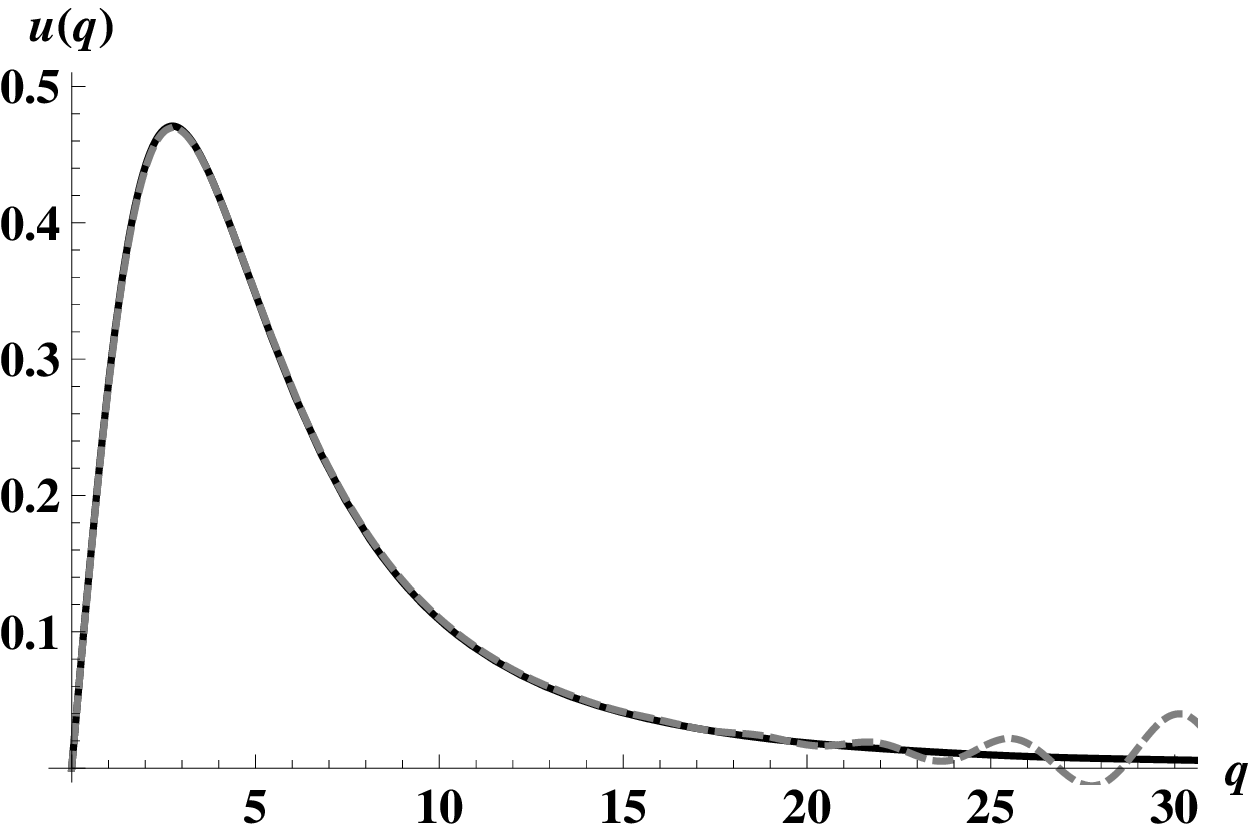}
\includegraphics*[width=0.45\textwidth]{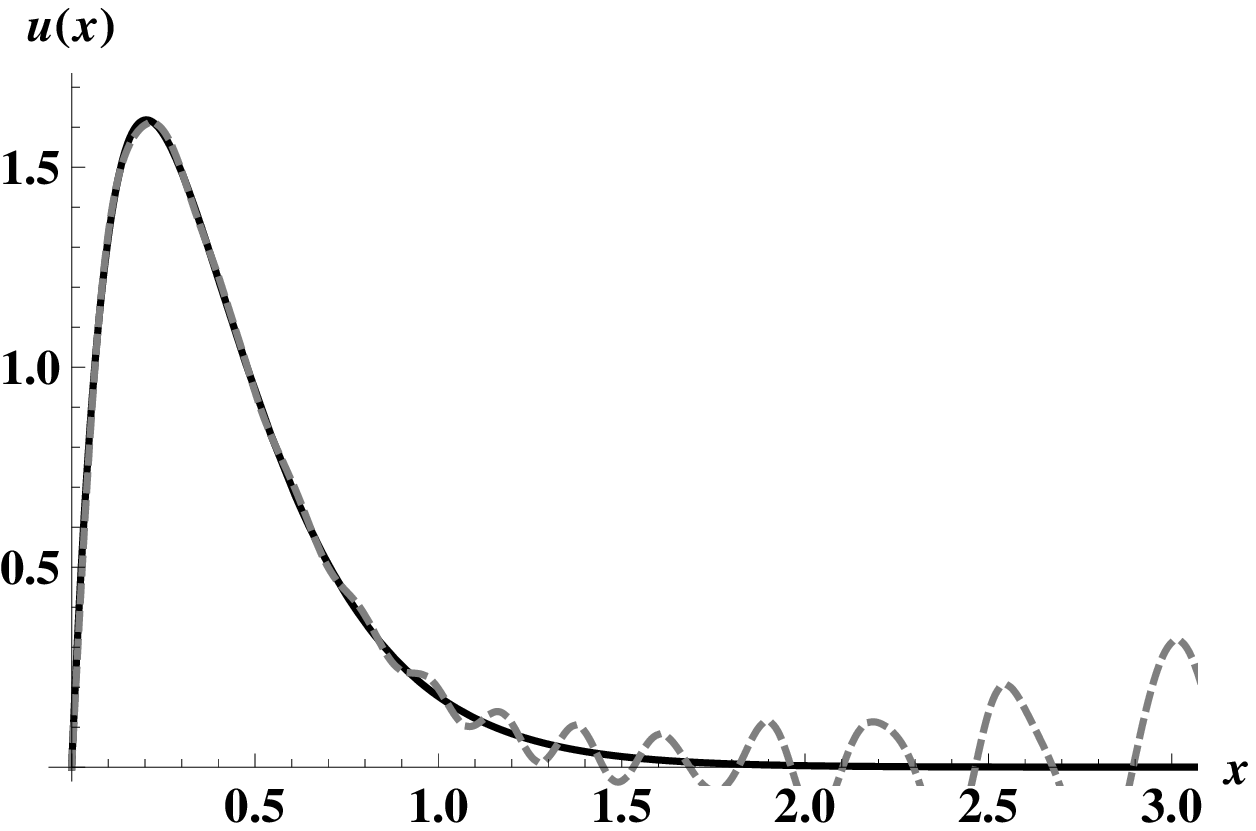}
\caption{Ground state of the dimensionless Hamiltonian (\ref{Hdim}) with $g=10$ (only positive values are shown). Left and right parts as in Fig.~\ref{fig:gausswf}. $h=0.5$, $N=20$ and $\epsilon=-16.2066$ ($h=0.05$, $N=20$ and $\epsilon=-16.3404$) for the computation in momentum (configuration) space.
\label{fig:yukwf}}
\end{figure}

The wavefunctions produced by the LMM in both spaces have also been compared for the Yukawa potential on Fig.~\ref{fig:yukwf}. With a small mesh of 20 points, the relative error on the ground state eigenvalues is around $10^{-10}$ for the computation in configuration space, and around $10^{-2}$ for the computation in momentum space. Curiously, the wavefunction computed in momentum space is already very well reproduced and is even better than the Fourier transform computed from the computation in configuration space with the same mesh. By increasing $N$, the beginning of the unphysical oscillations in the Fourier transforms can be rejected far in the asymptotic tail. 

\begin{figure}[htb]
\includegraphics*[width=0.45\textwidth]{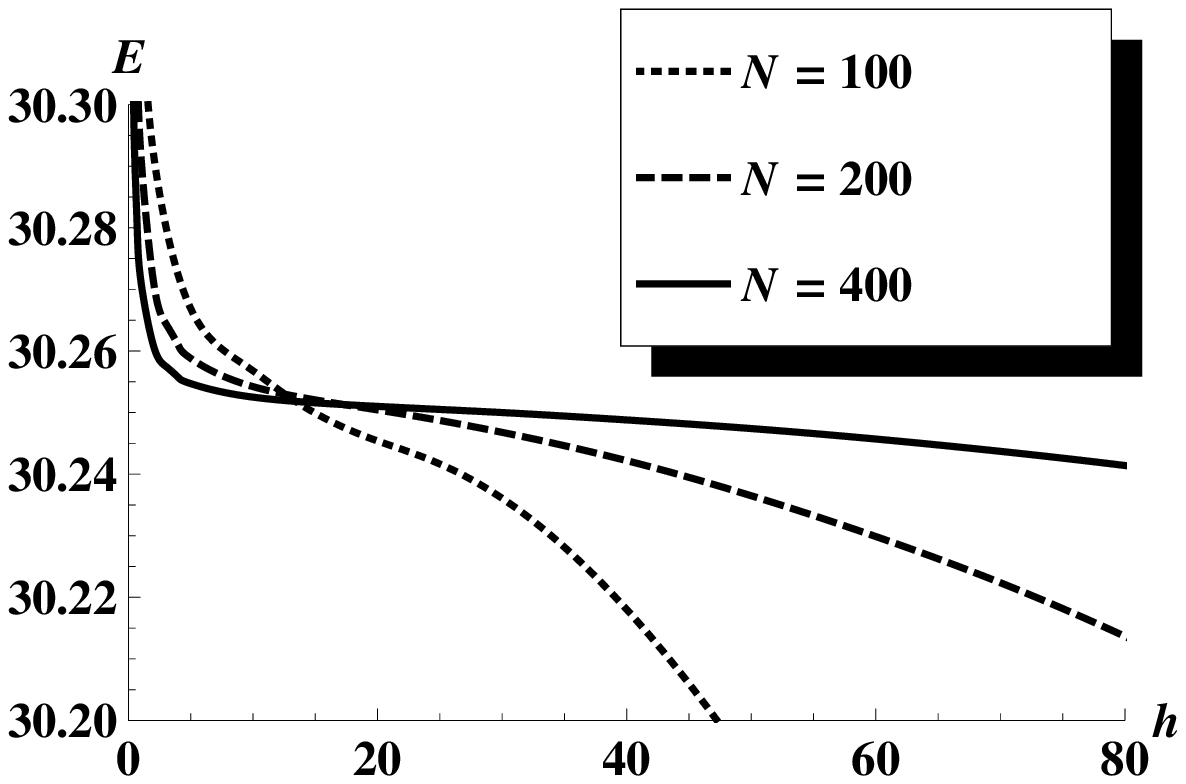}
\includegraphics*[width=0.45\textwidth]{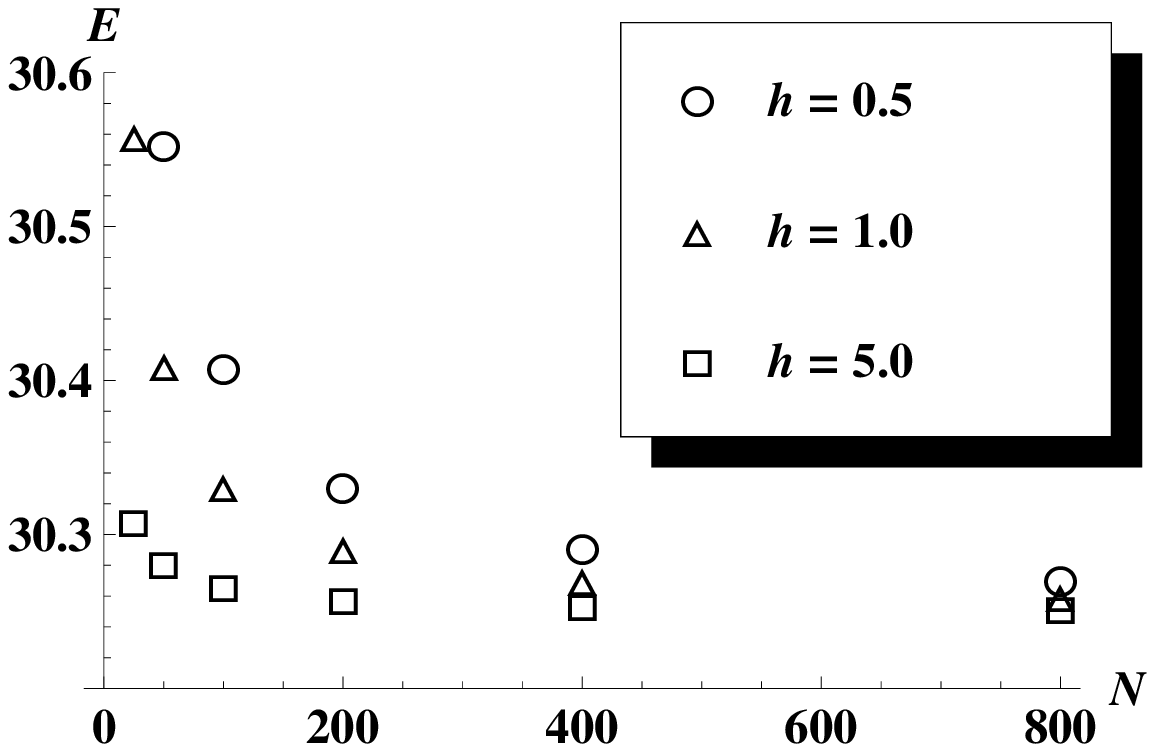}
\caption{Ground state eigenvalue $E$ of (\ref{Pexp4}) for the spinless Salpeter Hamiltonian with the Yukawa interaction considered (\ref{param}), as a function of $h$ and $N$. For $N=25$ and $h=0.5$, the value of $E$ is 30.81 far above the range of the graph.
\label{fig:SRYuk}}
\end{figure}

We will briefly comment the Yukawa interaction with a semirelativistic kinematics. As in the nonrelativistic case, we did not try to study a ``realistic" system. Quite arbitrary but convenient values are considered for the parameters,
\begin{equation}
\label{param}
m_1=m_2=m=16,\quad a=1, \quad b=5, 
\end{equation}
for which only one bound state exists ($0 < E < 2 m$). Results about the convergence are presented in Fig.~\ref{fig:SRYuk}. Graphics are similar to the ones produced in the nonrelativistic case: Existence of plateaus for energy as a function of $h$; increase of the length of the plateau with $N$. But, the convergence is slower than in the nonrelativistic computations: $N$ must be greater to achieve a quasi flat plateau (Fig.~\ref{fig:SRYuk}, left) and to reach a convergence between eigenvalues computed with different values of $h$ (Fig.~\ref{fig:SRYuk}, right). We have checked that it is also the case to obtain an agreement between observables computed in both spaces and to obtain better Fourier transforms. With a semirelativistic Hamiltonian, the kinetic energy increases as $\sqrt{\vec p\,^2}$ not as $\vec p\,^2$. So, higher values of the momentum can be reached by the wavefunction, and it can be more extended than with a nonrelativistic kinematics. This amplifies the problem mentioned in the previous section about the asymptotic behavior of the potential. 

\section{Concluding remarks}
\label{sec:conclu}

The Lagrange-mesh method is a procedure to compute accurate eigenvalues and eigenfunctions of quantum equations. Implemented at the origin in the configuration space, this technique requires only the computation of the potential at some mesh points \cite{BH-86,sema01}. This is due to the use of a Gauss quadrature rule with the fact that the basis functions satisfy the Lagrange conditions, that is to say they vanish at all mesh points except one. 

Using this very special property, we have shown that the Lagrange-mesh method can be adapted to be used directly in momentum space with a nonlocal potential (but local in the configuration space). In this case, nonrelativistic and semirelativistic kinetic parts are treated with the same manner. Moreover, it is possible to treat interactions which present discontinuities in configuration space. The wavefunction in momentum space is directly obtained under the form of expansion coefficients for the Lagrange functions. Using only these coefficients, it is easy to compute the wavefunction in configuration space by Fourier transform, as well as mean values of $p$-dependent or $r$-dependent operators. Though the technique is not variational, the convergence can be checked: Eigenvalues present a plateau as a function of the sole nonlinear parameter of the method; eigenvalues computed with different nonlinear parameters in a plateau tends towards the same value as the number of mesh points increases. 

The method has been tested with two potentials. With a Gaussian interaction and a nonrelativistic kinematics, a good accuracy can be obtained for eigenvalues and observables with a very small mesh, as for the method in configuration space: 20 points are probably sufficient for most of the physical applications. For a Yukawa interaction, it is necessary to use more points than for the Lagrange-mesh method in the configuration space, typically 10 times more. For both potentials: \emph{i}) The convergence is slowed down for a semirelativistic kinematics, as for the method in configuration space. But a good accuracy can nevertheless be reached \cite{sema01}; \emph{ii}) It is necessary to use a large mesh to obtain a correct Fourier transform of the wavefunctions, as for the computations in configuration space \cite{lacr11}. \emph{iii}) A good wavefunction in momentum space can obtained with a small mesh by a direct computation in momentum space. It does not present the unphysical oscillations of the Fourier transform of the wavefunctions in configuration space, which can only be eliminated by using a large mesh.

The purpose of this paper is to test the feasibility and the reliability of the Lagrange-mesh method in momentum space. We think that it can be safely applied to physical problems, since convergence tests exists and a good accuracy can always be obtained. Even if several hundreds of points are necessary to treat some potentials, eigenvalues can be rapidly computed. The method is very easy to implement once the interaction is known in momentum space, and it can be particularly useful if the kinetic part $T(\vec p\,^2)$ is an unusual function of the momentum, since its corresponding matrix is diagonal and easy to compute. This kind of situation appears in hadronic physics where quarks or gluons can be considered with a momentum dependent mass which can be very complicated to define in the configuration space \cite{szcz96,llan00,agui11}, e.g. $T(\vec p\,^2)=\sqrt{\vec p\,^2+m^2(\vec p\,^2)}$. For these systems, the dominant interaction can be simulated by a linear potential in the configuration space. This type of potential is \textit{a priori} highly singular in the momentum space, but it exists in a distributional sense and can be used in this context \cite{hers93}. 

\section*{Acknowledgments}

G.L. and C.S. would thank the F.R.S.-FNRS for the financial support.

\end{document}